\documentclass{jfm}

\usepackage{array}
\usepackage{amsmath,amssymb}
\usepackage{graphicx}
\usepackage{natbib}
\usepackage{multirow}
\usepackage{gensymb}
\usepackage{hhline}
\usepackage{hyperref}
\usepackage{breakcites}

\usepackage{color}
\begin{document}

\title{Spatiotemporal measurement of surfactant distribution on gravity-capillary waves}

\author{Stephen L. Strickland$^1$, Michael Shearer$^2$, Karen E. Daniels$^1$}

\affiliation{
$^1$Dept. of Physics, NC State University, Raleigh, NC, USA \\
$^2$Dept. of Mathematics, NC State University, Raleigh, NC, USA
}

\maketitle

\begin{abstract}
Materials adsorbed to the surface of a fluid -- for instance, crude oil, biogenic slicks, or industrial/medical surfactants -- will move in response to surface waves.
Due to the difficulty of non-invasive measurement of the spatial distribution of a molecular monolayer, little is known about the dynamics that couple the surface waves and the evolving density field.
Here, we report measurements of the spatiotemporal dynamics of the density field of an insoluble surfactant driven by gravity-capillary waves in a shallow cylindrical container.
Standing Faraday waves and traveling waves generated by the meniscus are superimposed to create a non-trivial surfactant density field. 
We measure both the height field of the surface using moir\'e-imaging, and the density field of the surfactant via the fluorescence of NBD-tagged phosphatidylcholine, a lipid.
Through phase-averaging stroboscopically-acquired images of the density field, we determine that the surfactant accumulates on the leading edge of the traveling meniscus waves and in the troughs of the standing Faraday waves. 
We fit the spatiotemporal variations in the two fields using an ansatz consisting of a superposition of Bessel functions, and report measurements of the wavenumbers and energy damping factors associated with the meniscus and Faraday waves, as well as the spatial and temporal phase shifts between them.
While these measurements are largely consistent for both types of waves and both fields, it is notable that the damping factors for height and surfactant in the meniscus waves do not agree, suggesting the presence of longitudinal waves.
\end{abstract}

\section{Introduction \label{s:Intro}}

The calming effect of surface oil on oceanic waves was recognised by sponge and pearl divers as early as the time of Pliny the Elder.
\citet{Franklin1774}, citing Pliny's account, also reported on sailors calming ocean waves with oil.
Franklin concluded from experiments that oil significantly damps short-wavelength, small-amplitude waves, and conjectured that it would also induce damping of larger long wavelength waves.
Lord Kelvin \citep{kelvin} first identified short-wavelength capillary-driven waves and long-wavelength gravity-driven waves as two regimes within the same mathematical formulation.
Kelvin's famous dispersion relation describing the wavelength-dependence of the speed of gravity-capillary waves shows a crossover from capillary to gravity waves at the minimum wave speed.

Much effort has since been devoted to studying the damping effect of surfactants on gravity-capillary waves \citep{Reynolds1880,Levich1941,Dorrestein1951,Goodrich1961a,Lucassen-Reynders1970}.
With or without a surfactant, energy is lost due to the viscosity of the fluid and the vorticity of the flow.
By adding a surfactant, the surface can experience additional tangential stresses which, balanced by viscous stress in the bulk, increase the vorticity of the bulk flow.
These tangential surface stresses are a consequence of the complex relation between surface stress and strain rate, which depends upon the bulk flow, surface velocity, and surfactant density.
The primary contribution to the tangential surface stress is the Marangoni effect \citep{Behroozi2007}, in which the advection of the surfactant monolayer by the passing wave results in gradients in the surfactant density field, inducing surface tension gradients and vorticity in the bulk fluid flow.
Thus, beyond quantifying the damping effect of surfactants on gravity-capillary waves, it is important to also understand the distribution of surfactant on the surface of the fluid.
In this paper, we combine two experimental techniques to measure both the surface height and surfactant density fields.
Using a physically-motivated ansatz, we quantify both fields and measure parameters such as the complex wavenumber and phase.
This provides a full spatiotemporal description, including the locations of surfactant accumulation within the wave pattern.

Surface waves can be generated and sustained in a variety of ways, depending on how energy is injected into the fluid system.
A vertically-vibrated fluid will generate both traveling meniscus waves, excited by the contact line at the container wall, and standing Faraday waves, first observed by \citet{Faraday1831}.
When the driving acceleration $a(t) = a_0 \sin(\omega t)$ is weak, only meniscus waves perturb the fluid surface. However, when $a_0$ is increased beyond a critical acceleration amplitude $a_c$, so that the injection of energy exceeds the dissipation due to the bulk viscosity, the meniscus wave becomes unstable to Faraday waves which grow to a finite amplitude \citep{Benjamin1954,chen1999}.
The parameter $a_c$ depends on the driving frequency $\omega$, the container geometry, and fluid properties \citep{Douady1990,Edwards1994,Bechhoefer1995,chen1999}.
The emergent Faraday waves can be either harmonic (frequency $\omega$) or subharmonic (frequency $\omega/2$), while the meniscus waves are always harmonic.

\begin{sloppypar}
Recently, there have been advances in our understanding of the relationship between the surface height field $h({\bf r},t)$ of meniscus waves and its interaction with surfactant molecules adsorbed to the surface, described by the density field $\Gamma({\bf r},t)$.
The theory of surfactant-laden meniscus waves in a cylindrical geometry, developed by \citet{Bock1991}, \citet{Saylor2000}, and \citet{Picard2006}, proceeds from the treatment of 2-dimensional traveling gravity-capillary waves in a Cartesian geometry \citep{Lucassen-Reynders1970}.
In this theory, the fluid motion in the incompressible bulk is modeled with the linearized Navier-Stokes equations, and the vertical and horizontal displacements of the surface satisfy the surface stress and kinematic boundary conditions. 
\citet{Bock1991} showed that for a fluid of equilibrium height $h_0$, the deviation in the surface height field $\Delta h({\bf r},t) = h({\bf r},t) - h_0$ for inward and outward traveling cylindrical gravity-capillary waves follow Hankel modes of the first and second kinds.
Recognizing that meniscus waves are a superposition of inward and outward traveling waves, \citet{Saylor2000} found that meniscus waves follow a $J_0$ Bessel mode.
Subsequently, \citet{Picard2006} derived an expression for the deviation of the surfactant density field from an equilibrium density $\Gamma_0$.
The quantity $\Delta \Gamma({\bf r},t) = \Gamma({\bf r},t) - \Gamma_0$ also follows a $J_0$ Bessel mode, with the same wavenumber and damping factor as the surface height field.
Because the motion of a fluid surface element depends upon the surface compression modulus $\epsilon = \Gamma_0 \frac{d \sigma}{d \Gamma}$, the magnitude of $\epsilon$ controls the phase shift between $\Delta h$ and $\Delta \Gamma$ fields \citep{Lucassen-Reynders1970}. 
\end{sloppypar}

For Faraday waves, \citet{Kumar2002,Kumar2004} used a linear stability analysis to determine the wave's critical acceleration and wavelength and predicted that the Faraday wave emerges with a spatial displacement (phase shift) between the fields $\Delta h$ and $\Delta \Gamma$.
Inspired by reports \citep{Douady1989} of a rotating Faraday wave, \citet{Martin2005} conjectured that the rotation was induced by the presence of a contaminant and showed that the drift can be generated when the reflection symmetry of the streaming flow breaks.
\citet{Ubal2005} executed numerical simulations of the gravity-modulated Navier-Stokes equations and predicted that neither $\Delta h$ nor $\Delta \Gamma$ evolve sinusoidally in time.
They found that $\Delta h$ lags behind $\Delta \Gamma$, characterised as a temporal phase shift between the two fields.

Many experiments with surfactant-laden gravity-capillary waves have focused on quantifying energy dissipation through either direct or indirect measurements of the surface height field alone \citep{Case1956, Davies1965, Jiang1993, Henderson1994, Henderson1998, Saylor2000, Behroozi2007}, and these works have demonstrated that the damping rate does not depend on whether the gravity-capillary waves are either standing or traveling.
Despite this progress, a quantitative understanding of the spatiotemporal dynamics of $\Gamma({\bf r},t)$ remains elusive.
Previous measurements of $\Gamma$ have utilized surface potential measurements in a small region near an electric probe \citep{Lange1984,Huhnerfuss1985b,Lange1986}.
These studies report a temporal phase shift between $\Delta \Gamma$ and $\Delta h$, with the caveat that it is necessary to account for both the response time of the probes and the finite time for the wave to pass from the surface potential probe to the wave height probe \citet{Lucassen-Reynders1987}.
In this paper, we present novel methods for measuring the spatial and temporal behaviors of $\Delta \Gamma$ and $\Delta h$, using stroboscopic imaging to avoid this complication.
These methods allow us to quantify the accumulation of surfactant relative to the crests/trough of the surface waves.
In addition, we test the extent to which the spatiotemporal dynamics can be described by a superposition of solutions for the meniscus and Faraday waves.

\begin{figure}
\centerline{\includegraphics[width=0.7\linewidth]{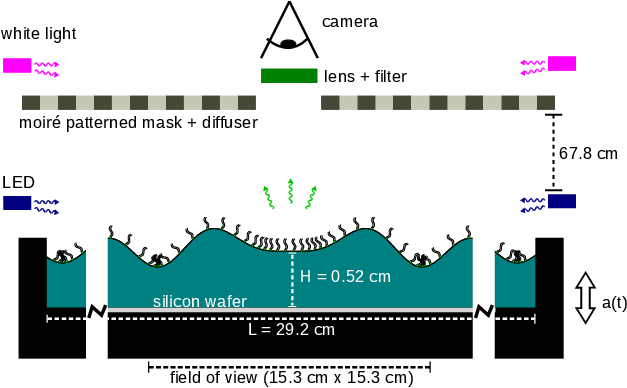}}
\caption{ (colour online)
{\bf Experimental apparatus.}
The container (black) is connected by low-friction air bushings to an electromagnetic shaker, and filled with a thin layer of water onto which a monolayer of NBD-PC surfactant is deposited.
The shaker vibrates the container vertically with acceleration $a(t) = a_0 \sin \omega t,$ generating meniscus and Faraday waves.
The fluorescence imaging system involves a ring of blue LEDs around the outer edge that excites the fluorophore on the lipids; digital images of the fluorescence emission are collected through an optical notch filter.
A silicon wafer sits flush with the bottom surface of the container, and reflects the excess light from the LEDs away from the camera.
The moir\'e imaging system consists of a patterned light source shining downward on the fluid surface, and the camera which records the distorted reflection of the pattern.
}
\label{f:App}
\end{figure}

Our experimental apparatus, shown schematically in figure~\ref{f:App} and described in detail in \S\ref{s:meth}, consists of a shallow cylindrical container holding a thin layer of water covered with a monolayer of fluorescently-tagged lipids.
We drive the system with vertical oscillations just above onset for the Faraday instability and make stroboscopic measurements of both the surface height field $h({\bf r},t)$ and the surfactant density field $\Gamma({\bf r},t)$. 
To obtain $h$, we illuminate the fluid with a target pattern, and use a combination of ray-tracing and nonlinear fitting to invert the resulting moir\'e image.
For $\Gamma$, we phase-average the fluorescence intensity from the tagged molecules using images acquired stroboscopically over many cycles of the driving oscillation, and convert these to quantitative measurements of the surfactant density field.
A key advantage of these techniques is that they are non-invasive.
Additionally, this and similar optical techniques \citep{Vogel2001,Fallest2010a,Strickland2014,Swanson2014} for measuring $\Gamma$ depend upon the mean distance between the fluorophores and therefore are not affected by the dynamics of either molecular rearrangement or domain formation/relaxation.

In \S\ref{s:results}, we show that our data are well-approximated by a linear superposition of a standing Faraday wave mode and a traveling meniscus wave mode.
We decompose the data into these two separate components in order to examine the spatiotemporal dynamics of each.
We determine that the surfactant accumulates on the leading edge of the meniscus waves and in the troughs of the Faraday waves. 
The meniscus waves are represented by $J_0$ Bessel functions.
The fields $\Delta h$ and $\Delta \Gamma$ have the same wavenumber within experimental error, but the damping factors for the two fields do not agree.
Since the temporal phase shift between the two fields is measured to be as large as 2 radians, in excess of the theoretical maximum of $\pi/2$ \citep{Lucassen-Reynders1970}, we conjecture that longitudinal waves \citep{Lucassen1968} may also be present. 
The Faraday waves are represented by $J_n$ Bessel functions. The fields $\Delta h$ and $\Delta \Gamma$ have the same symmetry number $n$ and the same wavenumber within experimental error. 
We observe that the whole Faraday wave pattern rotates around the center of the pattern in the same direction as an observed spatial phase shift between $\Delta h$ and $\Delta \Gamma$ fields. 
Both fields evolve sinusoidally and are temporally phase shifted by roughly $2.4$ radians. 

In \S\ref{s:Disc}, we discuss our observations about surfactant-covered meniscus and Faraday waves in context of their respective theoretical frameworks.
We also consider the possible presence of resonantly-excited longitudinal waves in our system, which could account for the anomalously large temporal phase shifts and the disagreement in the damping factors.
We highlight the suitability of using measurements of $\Delta \Gamma$ and the temporal phase shift between $\Delta h$ and $\Delta \Gamma$ as a way to probe the interfacial rheology.
Finally, we contrast the dynamics of these molecular monolayers with the dynamics of a monolayer of millimetric sized particles \citep{Sanl2014}.

\section{Experiment \label{s:meth}}

We excite traveling meniscus waves and standing Faraday waves on a surfactant-covered fluid layer by subjecting the system to a vertical sinusoidal oscillation, shown schematically in figure~\ref{f:App}.
A mechanical driving system provides the vertical acceleration, and two quasi-independent imaging systems measure the response of both the surfactant and the fluid surface.
The fluorescence imaging (FI) system measures the surfactant density field $\Gamma({\bf r},t)$ (details provided in \S\ref{s:FI}), and the moir\'e imaging (MI) system measures the surface height field (\S\ref{s:MI}).
Finally, we use a physically-motivated ansatz (\S\ref{s:Ansatz}) to support the ray-tracing and image analysis in order to convert the MI data to a surface height field $h({\bf r}, t)$ and to quantify properties of the deviations of the surface height field $\Delta h({\bf r},t) = h({\bf r},t) - h_0$ and surfactant density field $\Delta \Gamma({\bf r},t) = \Gamma({\bf r},t) - \Gamma_0$.

We perform all experiments just above the onset of the Faraday instability in order to excite a stable (non-chaotic) surface mode.
To reach this regime, we quasi-statically increase the driving acceleration in steps of $1.5\times10^{-3} g$ until the Faraday wave pattern is just identifiable in the moir\'e images.
Each quasi-static step during this preparation stage lasts for 90~s, during which the driving acceleration is held constant.
In the data-collection stage, the amplitude of the voltage signal for the electromagnetic shaker is held constant while we alternately collect MI and FI data. 
Although we collect data just above the Faraday wave onset, we nonetheless observe that the entire pattern rotates on the order of $10^{\circ}/$min.
Similar rotations have been observed in other experiments, with the rotation axis slightly displaced from the center of the container \citep{Gollub1983,Douady1989}.
During image analysis, we remove the rotation by rotating images back to a common reference.

\begin{figure}
\centering
\includegraphics[width=0.7\linewidth]{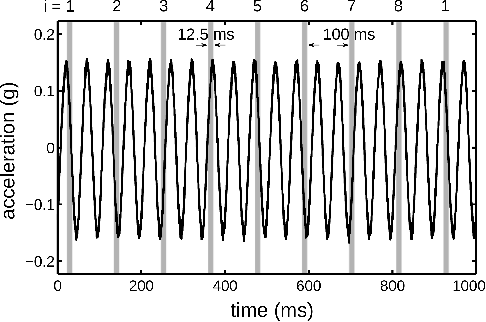}
\caption{ 
{\bf Stroboscopic imaging} allows for the collection and phase-averaging of low-intensity fluorescence images over many cycles.
During each second of the experiment, the container oscillates 20 times with an acceleration $a(t)$ (black curve).
During that second, the camera is triggered at four different phases of the oscillation (eight different phases of the Faraday wave oscillation).
The eight distinct exposures (gray bars) are spaced $1$ Faraday wave period apart with an exposure time of $1/8$ of a Faraday wave period, illustrated by the width of the gray bars. 
The same stroboscopic pattern was used for both experiments, and both the FI and MI data.
\label{f:CamTrig}}
\end{figure}

To increase the signal-to-noise ratio, all fluorescence imaging is performed stroboscopically to phase-average over the low-intensity signal.
The trigger pattern for the camera is chosen so that we cyclically sample eight different phases ($i = 1 \ldots 8$) of the Faraday waves (four phases of the meniscus waves), as shown in figure~\ref{f:CamTrig}.
In order to measure sub-period dynamics, we set the exposure time for each image to be $1/8$ of the subharmonic Faraday oscillation period, and average 700 stroboscopically-acquired images at each of the eight phases.
Obtaining this fluorescence composite data takes 14 minutes in total, during which time the Faraday wave rotates through roughly $140^\circ$.
We take a series of 32 moir\'e images during each minute of the data collection to account for the rotation and to measure the surface height field.

\subsection{Apparatus and Materials \label{s:Material}}

In order to produce stable Faraday waves, the vertical driving requires small amplitude steps during the preparation stage, small drift in the acceleration amplitude during the data-collection stage, and pure sinusoidal driving at all times. 
We achieve these requirements with a PM50A MB Dynamics shaker (peak force 220~N) mounted on a vibration isolation table.
The driving acceleration is transferred from the shaker to the container by two parallel shafts.
Each shaft is guided by a low-friction air bushing, and the parallel placement of the two bushings suppresses rotational motion.
The mounting bar for the container contains 2 two-axis ADXL203 accelerometers, sampled at a rate of $20$~kHz with a noise threshold of approximately $10^{-3} g$.
We achieve an uncertainty in $a_0$ of $5\times 10^{-3} g$ and all higher harmonics are at least a factor of $10^{-3}$ smaller than $a_0$.

The choice of materials in our experiments is guided by two important considerations: the sensitivity of surface experiments to preparation and handling, and the requirement of a low fluorescence background. 
The black-anodized container is made of aluminum machined into a cylindrical well of radius 14.6~cm. 
A 200~mm silicon wafer is embedded into the base of the well so that the wafer's top surface is flush with the bottom of the well. 
The silicon wafer provides both a reproducible substrate and excellent reflectivity \citep{Strickland2014}.

Prior to each experiment, we clean the container with detergent and perform a final rinse with 18.2 {M\ohm} water before drying with dry nitrogen gas.
The silicon wafer is cleaned for 5 minutes in an oxygen plasma environment, and all glassware is cleaned by soaking for several hours in a 2\% Contrad 70 solution.
These materials are rinsed with 18.2 {M\ohm} water and dried with dry nitrogen gas immediately before the experiment.
To initialize each experiment, we fill the container to a depth of $h_0 = 0.37 \pm 0.02$~cm (below the brim of the container) with 18.2~{M\ohm} water. 
Using a micropipette, we deposit a solution of chloroform and NBD-PC (1-palmitoyl-2-{12-[(7-nitro-2-1,3-benzoxadiazol-4-yl)amino]dodecanoyl}-sn-glycero-3-phosphocholine from Avanti Polar Lipids) onto the clean water surface. 
The concentration of NBD-PC in solution is 1 mg/mL, and due to the low interfacial tension between chloroform and water, the droplets spread over the fluid surface.

In this paper, we report on two experimental runs: Experiment 1 with $\Gamma_0 = 0.3 \, {\mu\text{g}}/{\text{cm}^2}$, and Experiment 2 with $\Gamma_0 = 0.2 \, {\mu \text{g}}/{\text{cm}^2}$.
Both runs are below the critical monolayer concentration for NBD-PC on water which is $\Gamma_c = 0.35 \, {\mu\text{g}}/{\text{cm}^2}$ \citep{Tsukanova2002}. 
The mean density $\Gamma_0$ is calculated from the deposited volume of chloroform solution and the known dimensions of the container. 
To reduce disturbance from external air currents and dust, the entire apparatus is enclosed in a plastic tent.
For Experiment 1, the temperature and humidity were $23.9 \pm 0.4 ^{\circ}$C and $18.5 \pm 0.5$\%, respectively; while for Experiment 2, they were $24.6 \pm 0.4 ^{\circ}$C and $18.0 \pm 0.5$\%.

\subsection{Fluorescence imaging \label{s:FI}}

The fluorescence imaging (FI) technique is based on quantifying spatiotemporal deviations of the surfactant density field ($\Delta \Gamma$) from the mean density by observing the fluorescence intensity in digital images \citep{Fallest2010a, Swanson2014, Strickland2014}. 
The NBD-PC surfactant molecule contains a fluorophore with an excitation peak at 464~nm and an emission peak at 531~nm.
Eight blue LEDs (1.5 W, 467 nm from Visual Communications Company, LLC) are mounted around the edge of the container to provide uniform excitation and illuminate the fluid surface at a low angle.
The silicon wafer substrate reflects unabsorbed light away from the imaging system, thus minimizing the background noise.
An Andor Luca-R camera fitted with a Newport $530\pm10$~nm bandpass filter is positioned above the center of the experiment and takes images of surfactant fluorescence within a field of view of width 15.4~cm.
Because the camera images its own reflection near the center of the system, we report measurements for an annular region of the fluid surface.

\begin{figure}
\centering
\includegraphics[width=0.7\linewidth]{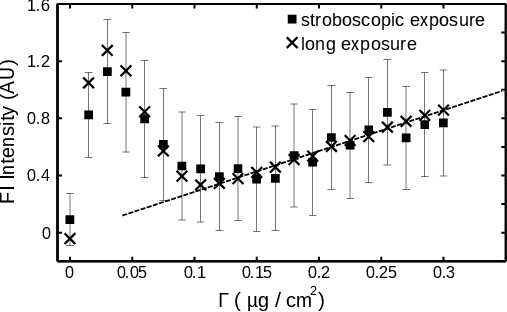}
\caption{ {\bf Calibration} of surfactant fluorescence intensity as a function of surfactant density.
The stroboscopic exposure data were obtained with the same exposure time, number of images, and gain settings as used for the FI technique.
The long exposure data was obtained with a single 2~s exposure and no gain, to confirm the trend.
The markers show the mean intensity of the pixels within the field of view while the error bars illustrate the standard error of the pixels for the stroboscopic exposure data.
The linear fit, shown as a dashed line, is used in the range $\Gamma > 0.1 \, {\mu \text{g}}/{\text{cm}^2}$ of the experiments.
}
\label{f:Calib}
\end{figure}

We calibrate the fluorescence intensity by performing experiments on a flat fluid surface covered with a known quantity of surfactant.
In order to ensure a uniform density on the surface, we deposit the chloroform-dispersed surfactant onto the clean surface of water, wait 60~s for the chloroform to evaporate, and then drive the system at 20~Hz and $0.2 g$ (above the Faraday onset) for 60~s.
This promotes redistribution of the surfactant, so that after turning off the shaker and waiting for the fluid to settle back to its flat state, the surfactant is uniformly distributed.
In this state, we collect two types of data: 700 images with the same exposure time as used for the FI composite images, and a final image with a single 2~s exposure time to provide a consistency-check on the trend.
The mean and standard error of the fluorescence intensity as a function of $\Gamma$, reported in figure~\ref{f:Calib}, are calculated from a composite of 700 images, as is done for each phase-averaged stroboscopically-acquired image.
We observe that the calibration is linear for $\Gamma > 0.1 {\mu \text{g}}/{\text{cm}^2}$.
This corresponds to a regime in which the equation of state relating surfactant density and surface tension is also linear \citep{Tsukanova2002}. 

\begin{figure}
\centering
\includegraphics[width=\linewidth]{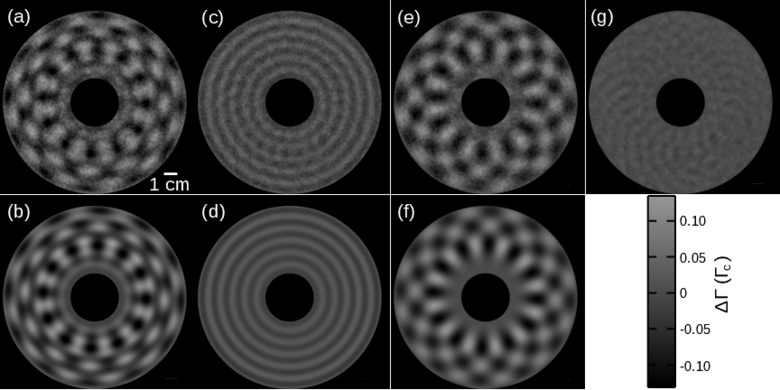}
\caption{ 
(colour online)
{\bf Sample FI composite image} (\emph{a}) showing $\Delta \Gamma$ from the $i=2$ stroboscopic phase of Experiment 1, smoothed by a Gaussian filter of $\sigma = 0.04$~cm.
(\emph{b}) The corresponding $\Delta \Gamma_T$, the best fit to \eqref{eqn:GProf}.
(\emph{c}) The meniscus wave (harmonic) component and (\emph{e}) the Faraday wave (subharmonic) component of the same $\Delta \Gamma$ shown in (\emph{a}).
These components are isolated by adding and subtracting (and dividing by $2$) the FI composite images from the $i=2$ and $i=6$ stroboscopic phases.
(\emph{d}) The meniscus wave (harmonic) component and (\emph{f}) the Faraday wave (subharmonic) component of the same $\Delta \Gamma_T$ shown in (\emph{b}).
(\emph{g}) This image is generated by subtracting (\emph{d}) from (\emph{c}).
The resulting pattern (smoothed with a Gaussian filter of $\sigma = 0.09$~cm) is a higher-order mode which is not accounted for in the ansatz \eqref{eqn:GProf}.
This mode has 24-fold symmetry and oscillates harmonically with peak-to-mean variation of $0.03 \, \Gamma_c$.
}
\label{f:SampleSurfData}
\end{figure}

To obtain the FI composite images we average 700 individual stroboscopically-acquired images taken at the same temporal phase (see figure~\ref{f:CamTrig}), each one corrected for the accumulated pattern rotation.
To measure $\Delta \Gamma$, we first subtract a background image (obtained by averaging all eight phases) and then use the linear fit shown in figure~\ref{f:Calib} to convert the light intensity values to local surfactant density values at each pixel.
A sample image is shown in figure~\ref{f:SampleSurfData} (\emph{a}), with the brighter areas corresponding to higher surfactant density.

\subsection{Moir\'e imaging \label{s:MI}}

\begin{figure}
\centering
\includegraphics[width=\linewidth]{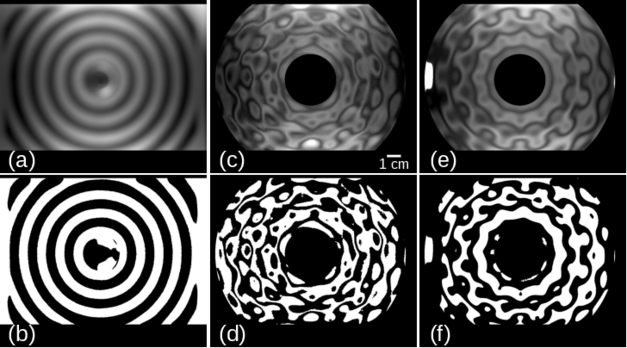}
\caption{
{\bf Sample MI images.}
(\emph{a}) Contrast-enhanced image and (\emph{b}) binarization of the target pattern reflected off a flat fluid surface. 
(\emph{c}) Contrast-enhanced moir\'e image and (\emph{d}) binarization of the target pattern reflected off the fluid surface at phase $i=2$ for Experiment 1. 
(\emph{e}) Ray-traced pseudo-image and (\emph{f}) binarization of the best-fit $h({\bf r})$ for image (\emph{c}) using parameters given in table~\ref{t:Params}.
}
\label{f:MISampleData}
\end{figure}

In order to measure the surface height field, we use a moir\'e imaging (MI) technique in which a reflected pattern of dark/light stripes reveals the spatial structure of the reflecting surface. 
For a known illumination pattern and known surface height field, a ray-tracing algorithm can produce a moir\'e pseudo-image which should correspond to the observed image.
To invert the moir\'e image, we assume a physically-motivated functional form for the surface height field (see \S\ref{s:Ansatz}) and search for the optimal parameters in that ansatz which best reproduce all eight observed images. 

The moir\'e imaging system consists of an Andor Luca-R camera which images the surface height field through a hole in the mask that creates the patterned light source (see figure~\ref{f:App}).
The light source pattern is a transparent film printed with a set of concentric black rings of width $1.27$~cm, chosen to be close to the wavelengths under consideration.
This film, as well as a green gel filter (Lee Filters 736 Twickenham Green, with a peak transmission wavelength of $525\pm50$~nm), are sandwiched between two panes of diffusing ground glass.
Two 500 W incandescent bulbs positioned above the pattern illuminate the fluid surface through this target pattern, and the camera records the light reflected off the fluid surface.
Even though the camera is focused on the fluid surface in order to optimize the low-intensity fluorescence measurements, the short focal length (35~mm) relative to the distance to the focal plane (67.8~cm) allows the use of a ray-tracing protocol to deduce the surface height profile.
Sample images of the patterned light source, MI images, and best-fit ray trace are shown in figure~\ref{f:MISampleData}.
To reduce the number of rays that trace back to either the camera or outside the patterned light source, we use the same annular field of view considered for the FI images.

Although the axes of the camera, pattern, and container are approximately aligned, we find it necessary to calculate their relative positions and orientations in order to achieve good ray-tracing results.
To obtain the camera position and orientation, we image a set of metal posts of known height and spacing; to measure the center of the target pattern, we image the reflection off of the flat fluid surface.
Using this data, we calculate the spatial resolution of the camera to be 0.0153~cm per pixel. 
To obtain the pixel coordinates of the system origin to sub-pixel resolution, we find the center of the meniscus wave pattern in composite data which has not been corrected for the Faraday wave rotation.
The pattern in the rotation-uncorrected composite data is nearly identical to that in figure~\ref{f:SampleSurfData} (\emph{c}).

To find the best fit parameters of the ansatz for $h({\bf r},t)$, we maximize the cross-correlation between the MI images and ray-traced pseudo-images. 
The ray-tracing protocol assumes that each ray corresponds to a pixel and all the rays start at a single point (the location of the camera)
The rays are assumed to reflect off the fluid surface at the $z=0$ plane, a justifiable assumption since the height of the wave peaks ($\mathcal{O}{\left(10 \mu \text{m} \right)}$) is much less than the distance between neighbouring peaks($\mathcal{O}{\left(1 \text{cm} \right)}$). 
When waves are present, the normal to the surface changes orientation and the reflected ray responds accordingly.
Each pixel of the ray-traced pseudo-images is assigned the intensity of the moir\'e pattern at the pixel nearest the intersection of the ray and the plane of the pattern.
For the few rays that trace back outside of the patterned light source, we assign the maximum intensity recordable by the camera.
We account for motion blur due to the oscillation of the wave by averaging three ray-tracing results for each of the eight MI images.

\subsection{Surface height field and surfactant density field ansatz \label{s:Ansatz}}

In order to compare $\Delta h$ and $\Delta \Gamma$, we use a physically-motivated ansatz which allows us to measure amplitudes, wavenumbers, and phases for the waves.
In addition, we assume that $\Delta h$ and $\Delta \Gamma$ are a linear superposition of the Faraday and the meniscus waves.
Our results will indicate that this assumption accounts for the most significant signals in the data. 
To distinguish the ansatz from the measured MI and FI data, we use the subscript $T$ for the ansatz quantities.

Meniscus waves take the form of a superposition of inward and outward traveling Hankel functions \citep{Saylor2000}.
The inward-propagating meniscus wave takes the form $\text{Re} \left[ e^{ i \omega t} H^{(1)}_0 (k r) \right]$ where the complex-valued wavenumber $k = k_{M} + i\alpha$ measures both the spatial wavenumber $k_M$ and the damping factor $\alpha$. 
Once at the center, the waves continue to propagate outward according to the form $\text{Re} \left[ e^{ i \omega t} H^{(2)}_0 (k r) \right]$.
The resulting superposition is $\text{Re} \left[ e^{ i \omega t} J_{0}(k r) \right]$.

\begin{sloppypar}
Sinusoidally-driven Faraday waves on an uncontaminated surface of an infinitely deep inviscid fluid \citep{Benjamin1954} or finite-depth low-viscosity fluid \citep{Edwards1994} take the form of a Bessel function $J_n(k_{F} r) \cos(n \theta) \cos( \frac{\omega}{2} t)$ where $\omega$ is the driving frequency and the spatial wavenumber $k_{F}$ is purely real.
We allow for the Bessel-mode to be displaced from the center of the container, to account for the observed rotational instability, and use cylindrical coordinates measured with respect to the pattern-center rather than the axis of the experiment. 
\end{sloppypar}

Assuming a linear superposition of the meniscus and Faraday waves, our ansatz for $\Delta h$ is given by
\begin{equation} \label{eqn:HProf}
\begin{split}
\Delta h_{T}(r,\theta,t) & = h(r,\theta,t) - h_0 = \Delta h_{MT}(r,t) + \Delta h_{FT}(r^\prime,\theta^\prime,t) \\
 & =
 A_{H} \text{Re} \left[ e^{ i ( \omega t + \phi_{HM} ) } J_{0}( (k_{HM} + i\alpha_{H}) r) \right] \\
 & + 
 B_{H} \cos( \frac{\omega}{2} t + \phi_{HF} ) J_{n}(k_{HF} r^\prime) \cos( n \theta^\prime + \delta_{H} ) \\
\end{split}
\end{equation}
where the subscripts $M,F$ indicate whether the parameter is associated with the meniscus wave or the Faraday wave, respectively.
The primed coordinates are relative to the center of the Faraday wave instead of the axis of the system.

\citet{Miles1967} predicts that $\Delta \Gamma_T$ of a gravity-capillary wave is proportional to the Laplacian of the velocity potential, which is in turn proportional to $\Delta h_T$ \citep{Lamb1879}.
Therefore, our $\Delta \Gamma$ ansatz takes the same form as \eqref{eqn:HProf}: 
\begin{equation} \label{eqn:GProf}
\begin{split}
\Delta \Gamma_{T}(r,\theta,t) = \Gamma(r,\theta,t) - \Gamma_0 & = \Delta \Gamma_{MT}(r,t) + \Delta \Gamma_{FT}(r^\prime,\theta^\prime,t) \\
& = 
 A_{\Gamma} \text{Re} \left[ e^{ i ( \omega t + \phi_{\Gamma M} ) } J_{0}( (k_{\Gamma M} + i\alpha_{\Gamma}) r) \right] \\
 & + 
 B_{\Gamma} \cos( \frac{\omega}{2} t + \phi_{\Gamma F} ) J_{n}(k_{\Gamma F} r^\prime) \cos( n \theta^\prime + \delta_{\Gamma} ) 
\end{split}
\end{equation}
Note that $\Delta \Gamma_{FT}$ can be separated into independent spatial and temporal factors.
In \S\ref{s:results}, we will test whether the temporal dynamics of $\Delta \Gamma_{F}$ follow the assumed $\cos( \frac{\omega}{2} t )$ behavior.

The nine fitting parameters in \eqref{eqn:HProf} (and the corresponding nine in \eqref{eqn:GProf}) each have a physical interpretation.
To distinguish the parameters of $\Delta h_T$ and $\Delta \Gamma_T$, we denote their parameters with the subscript $H,\Gamma$ respectively. 
The amplitudes of the meniscus and Faraday waves are given by $A_{H}$ and $B_{H}$, respectively.
The spatial wavenumbers of the pattern are: $k_{HM}$ and $k_{HF}$ (real, radial for both waves), $\alpha$ (imaginary, radial for meniscus wave), $n$ (real, azimuthal for Faraday wave).
The value of $\alpha$ corresponds to the energy damping rate of the meniscus wave.
Only the relative phase between $\Delta \Gamma$ and $\Delta h$ is independent; we report temporal phase shifts for both the meniscus wave ($\phi_{H M} - \phi_{\Gamma M}$) and the Faraday wave ($\phi_{H F} - \phi_{\Gamma F}$).
Finally, there is also a spatial phase shift for the Faraday waves ($\delta_H - \delta_\Gamma$). 

Due to the high-dimensionality of the parameter space, finding a best fit to this ansatz takes place in several stages, described below.
The full set of fitting parameters for both $\Delta h_T$ and $\Delta \Gamma_T$ are provided in table~\ref{t:Params}, and discussed in detail in \S\ref{s:results}.

First, we decompose the FI images into a harmonic component (to fit the meniscus-wave terms) and a sub-harmonic component (Faraday terms) (see figure~\ref{f:SampleSurfData} (\emph{c},\emph{e})).
This can be done by simply adding or subtracting images that are precisely one half Faraday wave period apart (e.g. images taken at phase $i=1$ added to/subtracted from images taken at $i=5$).
We use a Levenberg-Marquardt algorithm to fit the resulting meniscus-series and Faraday-series to the corresponding terms in \eqref{eqn:GProf}.
A sample comparison of the FI composite data and the best-fit $\Delta \Gamma_T$ are shown in figure~\ref{f:SampleSurfData} (\emph{b},\emph{d},\emph{f}).

Solving the inverse problem using the MI images is more challenging, and starts from an initial $\Delta h_T$ chosen using the fluorescence-determined wavenumbers (with the missing parameters chosen by hand).
Using MATLAB {\it fminsearch}, we simultaneously maximize the cross-correlation of all eight MI images against their corresponding ray-traced pseudo-images.
To minimize the effects of pattern noise on the cross-correlation, we binarize and then Gaussian blur both the MI image and the ray-traced pseudo-image before performing the cross-correlation.
The searching algorithm proceeds by varying the parameters in \eqref{eqn:HProf} until a best fit is found.
A sample image is shown in figure~\ref{f:MISampleData}, illustrating that we are able to fit even small features of the surface waves. 

\begin{table}
\centering

\begin{tabular}{ c | c | c | c | c }
 & & & Experiment 1 & Experiment 2 \\
\hhline{=|=|=|=|=}

 & fluid thickness & $h_0$ (cm) & $0.37$ & $0.37$ \\
\cline{1-5}
 & mean surfactant concentration & $\Gamma_0$ (${\mu \text{g}}/{\text{cm}^2}$) & $0.3$ & $0.2$ \\
\cline{1-5}
 & acceleration amplitude & $a_0$ (g) & $ 0.145 $ & $ 0.140 $ \\
\cline{1-5}
 & symmetry number & $n$						& $ 12 $					& $ 11 $ \\
\hhline{=|=|=|=|=}
\parbox[t]{2mm}{\multirow{5}{*}{\rotatebox[origin=c]{90}{ $\Delta \Gamma_T$ }}}
 & meniscus: amplitude & $A_{\Gamma}$ ($\Gamma_c$)							& $ 0.22 $		& $ 0.04 $ \\
 & meniscus: wavenumber & $k_{\Gamma M}$ (cm$^{-1}$)			& $ 6.70 $		& $ 6.56 $ \\
 & meniscus: damping & $\alpha_{\Gamma}$ (cm$^{-1}$)		& $ -0.15 $		& $ -0.30 $ \\
 & Faraday: amplitude & $B_{\Gamma}$ ($\Gamma_c$)							& $ 0.45 $		& $ 0.20 $ \\
 & Faraday: wavenumber & $k_{\Gamma F}$ (cm$^{-1}$)						& $ 3.40 $		& $ 3.24 $ \\
\hhline{=|=|=|=|=}
\parbox[t]{2mm}{\multirow{5}{*}{\rotatebox[origin=c]{90}{ $\Delta h_T$ }}}
 & meniscus: amplitude & $A_{H}$ (cm)										& $ 0.0022 $	& $ 0.0033 $ \\
 & meniscus: wavenumber & $k_{HM}$ (cm$^{-1}$)		& $ 6.6 $		& $ 6.5 $ \\
 & meniscus: damping & $\alpha_{H}$ (cm$^{-1}$)		& $ -0.17 $		& $ -0.11 $ \\
 & Faraday: amplitude & $B_{H}$ (cm)										& $ 0.021 $		& $ 0.021 $ \\
 & Faraday: wavenumber & $k_{HF}$ (cm$^{-1}$)								& $ 3.4 $		& $ 3.3 $ \\
\hhline{=|=|=|=|=}
 & meniscus: temporal phase shift & $ \phi_{H M} - \phi_{\Gamma M} $							& $ 1.19 $		& $ 2.03 $ \\
 & Faraday: temporal phase shift & $ \phi_{H F} - \phi_{\Gamma F} $							& $ 2.38 $		& $ 2.42 $ \\
 & Faraday: spatial phase shift & $ \delta_{H} - \delta_{\Gamma} $						& $ 0.33 $		& $ 0.76 $ 
\end{tabular}

\caption{
Summary of experimental parameters and best-fit parameters (\eqref{eqn:HProf} and \eqref{eqn:GProf}) for the two experiments.
The uncertainty of all parameter values is in the last reported digit.
The center of the Faraday wave (axis of patten rotation) is within $\pm 3.5$ mm of the container axis. 
}
\label{t:Params}
\end{table}

\section{Results \label{s:results}}

\begin{figure}
\centering
\includegraphics[width=\linewidth]{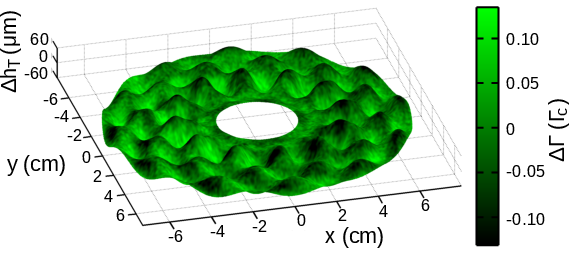}
\caption{
(colour online)
{\bf Visualization of $\Delta h_T$ and $\Delta \Gamma$} for Experiment 1, within the annular field of view.
The best fit using the $h$-ansatz \eqref{eqn:HProf} (mesh surface) and the FI composite data (colouration) for the $i=2$ stroboscopic phase.
The peak-to-mean variations in $\Delta h_T$ and $\Delta \Gamma_T$ are $\pm 60~\mu$m and $\pm 0.15~\Gamma_c$ respectively.
The $\Delta \Gamma$ fields are smoothed by a Gaussian filter of $\sigma = 0.04$~cm to reduce noise. 
In figures~\ref{f:MWDataCuts} and \ref{f:FWDataCuts}, this same data is decomposed into into its meniscus and Faraday wave components.
A video version of this plot is available in the online supplemental material.
}
\label{f:3DData}
\end{figure}

\begin{figure}
\centering
\includegraphics[width=\linewidth]{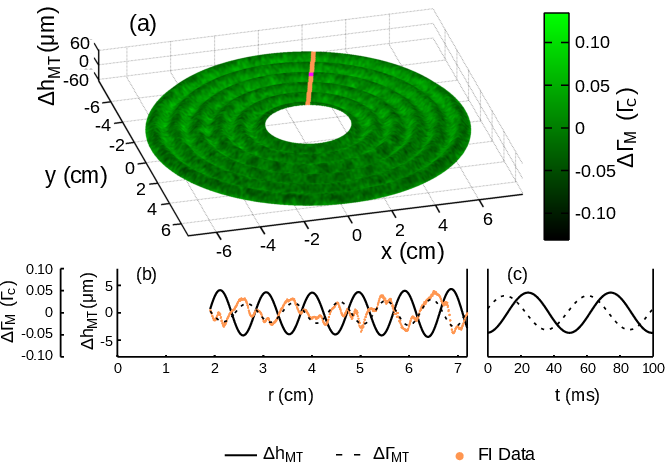}
\caption{
(colour online)
{\bf Meniscus wave} component of $\Delta h_T$ and $\Delta \Gamma$ for Experiment 1.
(\emph{a}) The best fit $\Delta h_{MT}$ ansatz from \eqref{eqn:HProf} (mesh surface) and the harmonic component of $\Delta \Gamma$ (colouration) for the $i=2$ stroboscopic phase.
(\emph{b}) Radial slice of both fields along $\theta = 2.83$~radians, the orange strip shown in (\emph{a}), at phase $i=2$.
The solid line is the best-fit height field $\Delta h_{MT}(r)$, and the surfactant density field is shown both as the data $\Delta \Gamma_{M}$ (orange points) and as the ansatz $\Delta \Gamma_{MT}$ (dashed line).
The lines are the Bessel modes given in \eqref{eqn:HProf} and \eqref{eqn:GProf}.
(\emph{c}) The temporal evolution of $\Delta h_{MT}$ (solid line) and $\Delta \Gamma_{MT}$ (dashed line) at the location $r = 4.79$~cm and $\theta = 2.83$~radians, marked by the magenta dot in (\emph{a}).
The lines are the sinusoids given in \eqref{eqn:HProf} and \eqref{eqn:GProf}. 
In all cases, the $\Delta \Gamma$ fields are smoothed by a Gaussian filter of $\sigma = 0.04$~cm to reduce noise. 
A video version of this plot is available in the online supplemental material.
}
\label{f:MWDataCuts}
\end{figure}

\begin{figure}
\centering
\includegraphics[width=\linewidth]{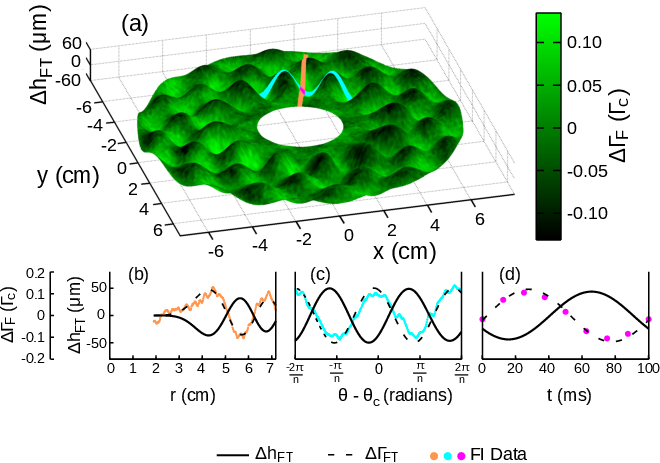}
\caption{
(colour online)
{\bf Faraday wave} component of $\Delta h_T$ and $\Delta \Gamma$ for Experiment 1.
(\emph{a}) The best fit $h_{FT}$ ansatz from \eqref{eqn:HProf} (mesh surface) and the sub-harmonic component of $\Delta \Gamma$ (colouration) for the $i=2$ stroboscopic phase.
(\emph{b}) Radial slice of both fields along $\theta_c = 2.83$~radians, the orange strip shown in (\emph{a}), at phase $i=2$.
The solid line is the best-fit height field $\Delta h_{FT}(r,\theta_c)$, and the surfactant density field is shown both as the data $\Delta \Gamma_{F}$ (orange points) and as the ansatz $\Delta \Gamma_{FT}$ (dashed line).
The lines are the Bessel modes given in \eqref{eqn:HProf} and \eqref{eqn:GProf}.
(\emph{c}) Azimuthal slice of both fields along $r_c = 4.26$~cm, the cyan strip shown in (\emph{a}), at phase $i=2$.
Here, the surfactant density field data $\Delta \Gamma_{F}$ is represented with cyan points.
The lines are the sinusoids given in \eqref{eqn:HProf} and \eqref{eqn:GProf} and plotted within a restricted range of the azimuthal dimension$\left[ -\frac{2 \pi}{n} , \frac{2 \pi}{n} \right] $ centered about $\theta_c$.
(\emph{d}) The temporal evolution of $\Delta h_{FT}$ (solid line) and $\Delta \Gamma_{FT}$ (dashed line) at the location $r_c$~cm and $\theta_c$~radians, marked by the magenta dot in (\emph{a}).
The lines are the sinusoids given in \eqref{eqn:HProf} and \eqref{eqn:GProf}. 
The magenta circles in (\emph{d}) plot behavior of $\mathcal{T}(t)$ given in \eqref{eqn:timeratio}, with errorbars much smaller than the symbol size. 
The $\Delta \Gamma$ fields in (\emph{a},\emph{b},\emph{c}) are smoothed by a Gaussian filter of $\sigma = 0.04$~cm to reduce noise. 
A video version of this plot is available in the online supplemental material.
}
\label{f:FWDataCuts}
\end{figure}

We present direct measurements of the spatiotemporal dynamics of a surfactant monolayer on gravity-capillary waves. 
In figures~\ref{f:SampleSurfData} and \ref{f:MISampleData}, we showed that our measurements of deviations of the surface height field $\Delta h$ and surfactant density field $\Delta \Gamma$ are well-approximated by a linear superposition of a Faraday wave and a meniscus wave, each represented by a Bessel function.
\eqref{eqn:HProf} and \eqref{eqn:GProf}, together with the parameters provided by fits to the data (see table~\ref{t:Params}), allow us to examine the spatiotemporal dynamics in detail. 

In figures~\ref{f:3DData}--\ref{f:FWDataCuts}, we present three-dimensional visualizations of the complete $\Delta h$ and $\Delta \Gamma$ fields (a snapshot at a single phase), as well as the same data split into its meniscus and Faraday wave components.
In each case, the best-fit $\Delta h_T$ is shown as a surface mesh, with the colouring representing the measured $\Delta \Gamma$: bright green regions are rich in surfactant, while dark green are depleted.
Movie versions of these figures, showing all eight phases, are available in the online supplemental material.
Below, we compare observations about both the traveling meniscus waves and the standing Faraday waves, along with the weaker higher-order mode observed in the FI composite data.

For the meniscus waves, we observe that the surfactant accumulates on the leading edge of the traveling wave, as shown in figure~\ref{f:MWDataCuts} (\emph{c}), where the dashed line ($\Delta \Gamma_{MT}$) leads the solid line ($\Delta h_{MT}$).
The fields $\Delta h$ and $\Delta \Gamma$ both exhibit a Bessel $J_0$ mode, and the real parts of the associated wavenumbers $k_{H M}$ and $k_{\Gamma M}$ agree, as expected.
We also observe the expected sinusoidal evolution predicted by \citet{Saylor2000} and \citet{Picard2006}. 
The maximum peak-to-mean variations in $\Delta h_{MT}$, measured within our annular field of view, are 5 and 8 $\mu$m for Experiments 1 and 2, respectively.
The corresponding values for $\Delta \Gamma_{MT}$ are 0.05 and 0.02 $\Gamma_c$.
These variations in $\Delta \Gamma_{MT}$ correspond to variations in the surface tension of 0.5 and 1.5 mN/m respectively \citep{Tsukanova2002}.
The location of the surfactant relative to the fluid wave corresponds to a temporal phase shift between $\Delta h_{MT}$ and $\Delta \Gamma_{MT}$, shown in figure~\ref{f:MWDataCuts} (\emph{c}).

For the Faraday waves, we observe that the surfactant accumulates in the troughs of the standing wave, as shown in figure~\ref{f:FWDataCuts} (\emph{d}), where the dashed line ($\Delta \Gamma_{FT}$) leads the solid line ($\Delta h_{FT}$). 
The fields $\Delta h$ and $\Delta \Gamma$ both exhibit a Bessel $J_n$ mode, and the wavenumbers $k_{H F}$ and $k_{\Gamma F}$ agree, as expected \citep{Miles1967}.
We note that the wavenumber of the Faraday wave is roughly double that for the meniscus wave, and the values of the wavenumbers can be captured by the finite depth Kelvin dispersion relation ($ \omega^2 = \left( g k + \sigma k^3 / \rho \right) \tanh(h_0 k) $) when the accepted values for the acceleration due to gravity $g$ and density of water $\rho$ are used and the same surface tension is left as a fitting parameter.
The best fit surface tensions for Experiments 1 and 2 are $33.3$ and $35.6$~mN/m respectively.
The fields $\Delta h$ and $\Delta \Gamma$ both exhibit a sinusoidal azimuthal behavior with the same symmetry $n$ as well as a sinusoidal evolution at half the driving frequency.
The maximum peak-to-mean variations in $\Delta h_{FT}$, measured within our annular field of view, are 62 and 64~$\mu$m for Experiments 1 and 2, respectively.
The corresponding values for $\Delta \Gamma_{FT}$ are 0.13 and 0.06~$\Gamma_c$.
These variations in $\Delta \Gamma_{FT}$ correspond to variations in the surface tension of 1.5 and 3.5 mN/m respectively \citep{Tsukanova2002}.
Similar magnitudes in the variations of the surface tension have been observed for $1$~Hz gravity waves contaminated with various molecular monolayers \citep{Lange1984}. 
The location of the surfactant relative to the fluid wave corresponds to a temporal phase shift between $\Delta h_{FT}$ and $\Delta \Gamma_{FT}$, shown in figure~\ref{f:FWDataCuts} (\emph{d}).
We also observe a small but non-negligible spatial phase shift ($\delta_{H} - \delta_{\Gamma}$) between $\Delta \Gamma_{FT}$ and $\Delta h_{FT}$ equal to 0.33 and 0.76 radians for Experiment 1 and 2 respectively.
This spatial phase shift is in the same direction as the rotation of the Faraday wave pattern in both experiments.

We have also considered a stricter test of the evolution for $\Delta \Gamma$ of the Faraday waves to determine whether they evolve non-sinusoidally, as indicated by the numerical solutions of \citet{Ubal2005}.
As given in \eqref{eqn:GProf}, the ansatz for $\Delta \Gamma_{F}$ can be factored into $\mathcal{S}_{T}(r',\theta') ~ \mathcal{T}_{T}(t)$ where the spatial factor is given by $\mathcal{S}_{T}(r',\theta') \equiv B_{\Gamma} ~ J_n(k_{\Gamma F} r') ~ \cos(n \theta' + \delta_{\Gamma})$ and the temporal factor by $\mathcal{T}_{T}(t) \equiv \cos(\frac{\omega}{2}t + \phi_{\Gamma F})$.
The subscript $T$ denotes these quantities are fits to the ansatz, rather than the data itself. 
Since the spatial factor is well-characterised by the ansatz ($\mathcal{S} = \mathcal{S}_T$), as shown in figures~\ref{f:SampleSurfData} (\emph{e},\emph{f}) and \ref{f:FWDataCuts} (\emph{b},\emph{c}), we can isolate the temporal evolution of $\Delta \Gamma_F$ for the data by considering the quantity 
\begin{equation}
 \mathcal{T}(t) = \left< \frac{\Delta \Gamma_{F}(r',\theta',t)}{\mathcal{S}_{T}(r',\theta')} \right>
 \label{eqn:timeratio}
\end{equation}
where the angle brackets denote an average over all space (approximately $5 \times 10^5$ pixels in total). 
In performing the average, we exclude pixels where $\left| \mathcal{S}_{T}(r',\theta') \right| < 0.006 ~ \Gamma_c$ to avoid the calculation being badly-conditioned. 
As shown in figure~\ref{f:FWDataCuts} (\emph{d}), the average temporal evolution $\mathcal{T}(t)$ 
 (and therefore the temporal evolution of $\Delta \Gamma_{F}$) is well-described by a sinusoidal function.

Our ansatz assumes a linear superposition of the meniscus and Faraday waves.
While this ansatz captures most of the total pattern, there is a residual mode in the FI data (see figure~\ref{f:SampleSurfData} (\emph{g})) which generates peak-to-mean variations in $\Delta \Gamma$ of $0.03$ and $0.01 \, \Gamma_c$ for Experiments 1 and 2 respectively.
This higher-order mode likely arises through a nonlinear interaction between the meniscus and Faraday waves.
Although this mode has a Bessel-mode shape like the Faraday waves ($F(r,\theta) \sim J_m(k r) ~ \cos(n \theta + \phi)$) with $m=2n$ (twice the symmetry number of the fundamental mode), the distance between neighbouring extrema is comparable to that of the meniscus waves.
Additionally, the higher mode is harmonic in time, unlike the subharmonic Faraday wave.

We observe that the degree of surfactant mobility depends on the type of wave.
By comparing the $\Gamma$-amplitudes of the associated Faraday vs meniscus Bessel modes $\left( \frac{B_\Gamma}{A_\Gamma} \right)$, we find a ratio of 2 for Experiment 1, and 5 for Experiment 2.
This indicates that Faraday waves have a larger effect than meniscus waves. 
However, if the $\Gamma$-amplitudes are rescaled by the associated $h$-amplitudes, then meniscus have proportional larger effect: the ratio $\left( \frac{B_\Gamma / B_H}{A_\Gamma / A_H} \right)$ is 0.21 for Experiment 1 and 0.78 for Experiment 2.
The cause of these differences is unclear and could be due to the difference between traveling vs. standing waves, to the difference in oscillation frequency, or to some subtlety of nonlinearity.
In contrast, the observed differences between the two experiments is clear and indicates that material mobility depends on $\Gamma_0$. 

There are other subtle differences between the two experiments, suggesting an array of rich dynamics that depend on the mean surfactant concentration.
First, the temporal phase shifts $\Delta \phi = \phi_H - \phi_\Gamma$ are consistent for the Faraday waves across both experiments, but differ for the meniscus waves.
Furthermore, only $\Delta \phi_M$ for Experiment 1 is less than $\pi/2$, the expected maximum temporal phase shift for linear gravity-capillary waves \citep{Lucassen-Reynders1970}.
Second, the two damping factors ($\alpha_H$ and $\alpha_\Gamma$, both associated with the meniscus wave) agree for Experiment 1 but not Experiment 2.
From the theoretical predictions of \citet{Picard2006}, we would expect agreement.

\section{Conclusions\label{s:Disc}}

We have successfully developed a novel technique for measuring the surface height $h$ and surfactant density $\Gamma$ fields for waves propagating on a surfactant-covered fluid. 
We observe that the surfactant accumulates on the leading edge of traveling meniscus waves and the troughs of standing Faraday waves.
The deviations of both the surface height field $\Delta h$ and surfactant density field $\Delta \Gamma$ take characteristic Bessel forms for both the meniscus and Faraday waves. 
For the meniscus waves, both fields follow a $J_0$ Bessel mode, but are temporally phase shifted from each other. 
For the Faraday waves, both fields follow the same $J_n$ Bessel mode, but are both temporally and spatially phase shifted from each other.

For meniscus waves, \citet{Saylor2000} and \citet{Picard2006} have analytically derived $\Delta h$ and $\Delta \Gamma$.
These two fields are expected to share the same functional form up to a complex valued coefficient ($F(r,t) = F_0 ~ J_0( k r ) ~ e^{i \omega t}$ where $k = k_{M} + i \alpha$) and therefore are expected to have the same wavenumbers and damping factors.
However, this work does not predict the relationship between the values of $F_0$ for the two fields, which would be necessary to understand the temporal phase shifts. 
\citet{Lucassen-Reynders1970} studied 2-dimensional traveling gravity-capillary waves in a Cartesian geometry, and derived a prediction for the phase shift between the surface area expansion and $\Delta h$.
For $\Gamma$ inversely proportional to the area of the fluid surface, the temporal phase shift between $\Delta h$ and $\Delta \Gamma$
is predicted to fall between $0$ and $\pi/2$ depending on the surface compression modulus.
This corresponds to the surfactant accumulating somewhere between the crests of the waves and the leading edge. 

In Experiment 2, we observe $\phi_{HM} - \phi_{\Gamma M} > \pi/2$ and $\alpha_H \neq \alpha_\Gamma$, both of which suggest that we may be exciting longitudinal waves in the system.
Longitudinal waves, like gravity-capillary waves, are a solution to the linearized Navier-Stokes equations for incompressible fluids.
By including the effect of the surfactant through the normal and tangential stress boundary condition, \citet{Lucassen1968b} derived a dispersion relation for which one branch corresponds to gravity-capillary waves, which have roughly equal parts transverse and tangential motion of the interface, and the other branch corresponds to longitudinal (i.e., Marangoni) waves, which have significantly more tangential motion than transverse \citep{Lucassen-Reynders1970}.
The possibility of a gravity-capillary wave resonantly exciting a longitudinal wave was first suggested by \citet{Lucassen1968b} because the maximum in energy dissipation coincides with the equality of the magnitude of the gravity-capillary and longitudinal wavenumbers.
Recently, \citet{Ermakov2003} proposed a mechanism for this resonant excitation which would explain the coincidence of damping with the wavenumber equality.
Future experiments which measure $\Delta \Gamma$ would be able to test this mechanism.

The discrepancies in the measured values of $\alpha$ highlight the importance of quantifying the surface dilational viscosity, dilational elasticity, and surface tension (collectively, the interfacial rheology).
Historically, techniques for measuring the interfacial rheology have relied on extracting the wavenumber and damping factor from the surface height field $h$ of a traveling gravity-capillary wave \citep{Miyano1983,Jiang1993,Saylor2000,Behroozi2007}.
The experiments described here provide a way to make a more direct measurement, using both the $h$ and $\Gamma$ fields.

For Faraday waves, there is theoretical literature relating the interaction between a surfactant monolayer and a Faraday wave in a fluid of arbitrary depth \citep{Kumar2002,Kumar2004,Martin2005,Ubal2005}.
\citet{Kumar2002,Kumar2004} assumed that in an infinitely broad fluid, both $\Delta h$ and $\Delta \Gamma$ oscillate sinusoidally and with zero temporal phase shift.
They also predicted a spatial phase shift between the two fields. 
In the experimental data, we observe sinusoidal oscillations, but with a temporal phase shift of approximately 2.4 radians. 
However, we confirm the spatial phase shift ($ \delta_{H} - \delta_{\Gamma} $ in table~\ref{t:Params}).
\citet{Martin2005} attribute rotational drift of Faraday waves to the generation of a streaming flow, under the assumption that the two fields have no spatial or temporal phase shift.
However, it is also plausible that our observed rotation of the Faraday waves is caused by the spatial phase shift.
\citet{Ubal2005} studied the motion of $\Delta h$ and $\Delta \Gamma$ through numerical simulations of the two dimensional Navier-Stokes equations with physically realistic interfaces and spatially symmetric boundary conditions in a finite depth fluid.
In agreement with our experiments, they predicted that $\Delta h$ and $\Delta \Gamma$ have similar spatial patterns, and that maxima in $\Delta \Gamma$ would precede maxima in $\Delta h$, an effect characterised by a temporal phase shift.
The prediction of peak-to-mean variations in $\Delta \Gamma \approx 0.03$ to $0.13 \Gamma_c$ is consistent with our observations, although for different parameters than used in the experiment.
However, their conclusion that the two fields have different non-sinusoidal time dependence does not agree with our data. 
Additionally, the use of spatially symmetric boundary conditions in the numerical simulations excludes the possibility of predicting the drift of the Faraday wave that we observe.

Finally, although we have studied the wave-driven accumulation of a molecular layer of surface contamination, similar effects are known to be present for other types of surface contamination.
For example, \citet{Sanl2014} used Faraday waves to redistribute a monolayer of millimeter-scale polystyrene spheres deposited on the surface.
Depending on the concentration of the particulate layer, the macroscopic contaminant would accumulate in either the anti-nodes (peak/trough) or the nodes of the waves.
This behavior is quite different from what we observe at the molecular scale, suggesting the presence of a crossover in particle size or surface activity.
Because oceanic contamination ranges in size from the molecular scale to flotsam and jetsam, more work is needed to develop a complete picture of the dynamics of contamination on surface waves.

\section{Acknowledgments}

This work was funded by the NSF, under grant number DMS-0968258. 
We are grateful to the West Virginia High Technology Consortium Foundation for the donation of the electromagnetic shaker and isolation table, and to Rachel Levy for the use of an Andor camera in the development of this work (funded by Research Corporation RCSA 19788).

\bibliographystyle{jfm}
\bibliography{StricklandReferencesV3}

\end{document}